\documentclass[12pt]{article}
\usepackage{graphicx,amsfonts,amsmath,amssymb,mathrsfs}

\title{Bell's Jump Process in Discrete Time}

\author{ 
Jonathan Barrett\footnote{Perimeter Institute for Theoretical Physics, 
    31 Caroline Street North, Waterloo, Ontario, Canada, N2L 2Y5.
    E-mail: jbarrett@perimeterinstitute.ca},
Matthew Leifer\footnote{Perimeter Institute for Theoretical Physics,
    31 Caroline Street North, Waterloo, Ontario, Canada, N2L 2Y5.
    E-mail: mleifer@perimeterinstitute.ca}, and
Roderich Tumulka\footnote{Mathematisches Institut,
    Eberhard-Karls-Unversit\"at, Auf der Morgenstelle 10, 72076
    T\"ubingen, Germany.  E-mail:
    tumulka@everest.mathematik.uni-tuebingen.de}
}
\date{September 27, 2005}

\newcommand{\Hilbert}{\mathscr{H}}
\newcommand{\conf}{\mathcal{Q}}
\newcommand{\Q}{\conf}

\renewcommand{\Re}{\mathrm{Re}}
\renewcommand{\Im}{\mathrm{Im}}

\newcommand{\PPP}{\mathbb{P}}
\newcommand{\RRR}{\mathbb{R}}
\newcommand{\CCC}{\mathbb{C}}

\newcommand{\ZZZ}{\mathbb{Z}}

\renewcommand{\sp}[2]{\langle #1|#2 \rangle}
\newcommand{\z}[1]{{#1}}

\begin{document}
\maketitle
\begin{abstract}
  The jump process introduced by J.~S.~Bell in 1986, for defining a
  quantum field theory without observers, presupposes that space is
  discrete whereas time is continuous. In this letter, our interest is to find 
  an analogous process in discrete time. We argue that a genuine 
  analog does not exist, but provide examples of processes in discrete time
  that could be used as a replacement.

\medskip

\end{abstract}

One of the central challenges for ``hidden variable'' approaches to
quantum mechanics, such as the de Broglie-Bohm pilot wave theory, is
to provide an adequate account of relativistic quantum field theory.
To address this, Bell introduced a jump process on a discrete lattice
\cite{Bell86, crea2A, crex1, schwer}, intended to reproduce the
quantum mechanical predictions for fermion number density in space.
The same method can be used to generate stochastic trajectories for
any discrete observable, both in field theory and in nonrelativistic
quantum mechanics.  For a discretized position observable in
nonrelativistic quantum mechanics, Bell's process reduces to the de
Broglie-Bohm pilot wave theory in the continuum limit \cite{Vink,
  Sudbery}, so it is a natural analog of this theory for discrete
``beables'' \cite{BellBeable}.

Although the ``beables'' in Bell's process are discrete, it still
contains a continuous time parameter.  However, there are several
reasons for developing a discrete-time version of the process.
Firstly, some approaches to quantum gravity are based on fundamentally
discrete space-time structures, so a realist account of these theories
along Bohmian lines would have to be fully discrete.  Secondly,
``hidden variable'' theories, no matter whether they are realized in
nature or not, can be useful for numerical simulations \cite{Lopreore,
  Dennis2}, visualizations \cite{Dennis, Struyve}, bookkeeping
\cite{Dennis}, and obtaining better intuitions about quantum
phenomena.  Numerical simulations are discrete by nature, and a fully
discrete theory may also be useful when dealing with quantum phenomena
usually described in a discrete setting, such as those considered in
quantum information and computation.  Thirdly, Valentini
\cite{Valentini} has recently proposed that matter in quantum
nonequilibrium, i.e. beables with distributions other than $|\Psi|^2$,
if existant, may provide astonishing computational resources, enabling
us to solve NP-complete problems in polynomial time.  However, since
classical analog computers can also outperform Turing machines if the
continuous variables can be manipulated with perfect accuracy, this
claim would be simpler to verify in a fully discrete model.

In this letter, we highlight the difficulties inherent in discretizing
Bell's jump process, and propose two concrete discretized processes
that circumvent them and converge to Bell's process as the time step
$\tau$ goes to zero.  Other possibilities exist, along the lines of
recent proposals by Aaronson \cite{Aar04}, and these will be developed
in future work.

Bell's process is a Markovian pure jump process $(Q_t)_{t \in \RRR}$
on a lattice $\Q$ with rate for the jump $q' \to q$ given by
\begin{equation}\label{crate}
  \sigma_t(q|q') = \frac{\bigl[ \tfrac{2}{\hbar}\, \Im \,
  \sp{\Psi_t}{P(q)HP(q')|\Psi_t} \bigr]^+}{ \sp{\Psi_t}{P(q')|\Psi_t} } \,,
\end{equation}
where $x^+ = \max(x,0)$ denotes the positive part of $x \in \RRR$,
$\Psi_t$ is the state vector of a quantum (field) theory, evolving in
some Hilbert space $\Hilbert$ according to
\begin{equation}\label{cschr}
  i\hbar \frac{d\Psi_t}{dt} = H\Psi_t \,, 
\end{equation}
$H$ is the Hamiltonian, and $P(q)$ is the projection to the subspace
$\Hilbert_q \subseteq \Hilbert$, where the $\Hilbert_q$ form an
orthogonal decomposition, $\Hilbert = \bigoplus_{q \in \Q} \Hilbert_q$.
Relevant properties of Bell's process are that at every time $t$, the
distribution of $Q_t$ is the quantum distribution 
\begin{equation}\label{Born}
  \sp{\Psi_t} {P(q)| \Psi_t} \,, 
\end{equation}
and that its net probability current between $q'$ and $q$, 
$\sigma_t(q|q') \PPP(Q_t = q') - \sigma_t(q'|q) \PPP(Q_t=q)$ 
\z{where $\PPP$ denotes ``probability,''} agrees with the
quantum expression for the probability current,
\begin{equation}\label{ccurrent}
  \tfrac{2}{\hbar}\, \Im \, \sp{\Psi} {P(q)HP(q')| \Psi} \,.
\end{equation}

Since many constructions are easier in discrete time than in
continuous time, one might have expected that there is an analogous
Markov chain $(\tilde Q_t)_{t \in \tau\ZZZ}$ on $\Q$ with discrete
time step $\tau$ such that the probability $\PPP_t(q' \to q)$ for the
transition $q' \to q$, i.e., the conditional probability $\PPP
(\tilde{Q}_{t+\tau} =q| \tilde{Q}_t = q')$, is given by a formula
similar to \eqref{crate}, with $H$ replaced by a simple function of
the unitary $U$ defining the time evolution
\begin{equation}\label{dschr}
  \Psi_{t+\tau} = U \Psi_t\,,
\end{equation}
and that one could arrive at this formula by
a reasoning similar to the one leading to \eqref{crate} \z{from 
\eqref{Born} and \eqref{ccurrent}, as} given in
\cite[Sec.~2.5]{crea2A}. 

However, this is not possible in any obvious way. The \z{obstacle} is
that in the time-discrete case there is no obvious formula for the net
probability current $J(q,q')$ between $q'$ and $q$, replacing \eqref{ccurrent}
of the continuous case. \z{Given an expression for $J(q,q')$ in terms 
of $\Psi,P$, and $U$, we could set}
\begin{equation}\label{transJ}
  \PPP_t(q' \to q) = \frac{J_t(q,q')^+} {\sp{\Psi_t}{P(q')|\Psi_t} }
  \quad \text{ for }q \neq q' \,,
\end{equation}
\z{which would define a Markov chain $(\tilde{Q}_t)_{t\in \tau\ZZZ}$ 
whose probability current} 
\begin{equation}\label{dcurrent}
  \PPP_t(q' \to q) \, \PPP(\tilde{Q}_t = q') - \PPP_t(q\to q') \, \PPP(\tilde{Q}_t = q)
\end{equation}
\z{coincides with $J(q,q')$ and whose distribution at any time $t$ coincides with the quantum distribution \eqref{Born}, provided $J(q,q')$ has the following properties: }
\begin{subequations}\label{cond}
\begin{align}
  J(q,q') &\in \RRR \label{cond1}\\
  J(q',q) &= - J(q,q') \label{cond2}\\
  \sum_{q\in \Q} J(q,q')^+ &\leq \sp{\Psi}{P(q')|\Psi} \label{cond3}\\
  \sum_{q' \in \Q} J(q,q') &= \sp{\Psi}{U^*P(q)U|\Psi} - \sp{\Psi}{P(q)|\Psi} \,. 
  \label{cond4}
\end{align}
\end{subequations}
\z{Currents of the form \eqref{dcurrent}, with transition probabilities \eqref{transJ} and distribution \eqref{Born}, have these properties by construction. Property \eqref{cond3} expresses that no greater amount of probability can get transported away from $q'$ than present at $q'$, and \eqref{cond4} guarantees the quantum distribution \eqref{Born} at the next time step. The obvious way of guessing a formula for $J(q,q')$ is to start from one of the expressions}
\begin{subequations}
\begin{align}
  &\sp{\Psi} {U^* P(q)UP(q')| \Psi}  \label{guess1}\\
  &\sp{\Psi} {P(q)UP(q')| \Psi} \,, \label{guess2}
\end{align}
\end{subequations}
\z{to multiply it by any numerical constant, to take the real or imaginary parts to ensure \eqref{cond1}, and to anti-symmetrize in $q$ and $q'$ to ensure \eqref{cond2}. However, all expressions thus obtained generically violate \eqref{cond4}, except for the anti-symmetrization of $2 \Re$\eqref{guess1},} 
\begin{multline}\label{expression}
  J(q,q') = \tfrac{1}{2} \sp{\Psi}{ \bigl( U^* P(q) UP(q') + P(q')U^*P(q)U -\\ 
  U^*P(q')UP(q) - P(q)U^* P(q')U \bigr) |\Psi} \,,
\end{multline}
\z{which can violate \eqref{cond3} 
  (numerically we found 46 examples of such violations among
  one thousand randomly chosen $U$ and $\psi$ in $\Hilbert = \CCC^3$
  with fixed one-dimensional projections $P(q)$ and $P(q')$).}

However, a different reasoning leads to \z{a process in discrete 
time that has some features in common with Bell's process.} 
Choose $H$ such that
\begin{equation}\label{UH}
  U = e^{-i\tau H} \,,
\end{equation}
so that the evolution \eqref{cschr} generated by $H$ is a continuation
of the evolution \eqref{dschr} generated by $U$. (The degree of
non-uniqueness of this choice is discussed later.)  Then, consider
Bell's process $(Q_t)_{t \in \RRR}$ in continuous time for this $H$.
By restriction to just the integer times, we obtain a Markov process
$\tilde{Q}_t := Q_t$ for $t \in \tau\ZZZ$.

The process $(\tilde{Q}_t)_{t \in \tau\ZZZ}$ has the quantum
distribution \eqref{Born} at every time. It is important for this that
the two evolution laws \eqref{cschr} and \eqref{dschr} for $\Psi$ lead
to the same $\Psi_t$ at every $t$ that is an integer multiple of
$\tau$.  It makes no sense to ask whether the probability current of
this process, $\PPP ( \tilde{Q}_{t + \tau} = q, \tilde{Q}_t = q') -
\PPP (\tilde{Q}_{t + \tau} = q', \tilde{Q}_t = q)$, agrees with the
one prescribed by quantum theory, since, as discussed above, quantum
theory does not prescribe a unique current in the discrete-time case.
Note that in the limit $\tau \to 0$ the process approaches Bell's
process. This fact and the simple and straightforward construction of
$(\tilde{Q}_t)_{t \in \tau\ZZZ}$ \z{suggest that this may be the closest 
one can get to an} analog of Bell's process in the time-discrete case.

The transition probability $\PPP_t (q' \to q) = \PPP
(\tilde{Q}_{t+\tau} =q| \tilde{Q}_t = q')$ does not, \z{however,} possess a simple
formula in terms of $\Psi_t$, $U$, and $P(\cdot)$ analogous to
\eqref{crate}, only the following one:
\begin{equation}\label{drate}
\begin{split}
  & \PPP_{t_0} ( q' \to q) =
  \sum_{n=0}^\infty \sum_{q_0, \ldots, q_n \in \Q} \delta_{q',q_0} \,
  \delta_{q,q_n} \int\limits_{t_0}^{t_0+\tau} dt_1 \int\limits_{t_1}^{t_0+\tau}
   dt_2 \cdots \int\limits_{t_{n-1}}^{t_0+\tau} dt_n \, \times \\
  &\quad \times \: \exp \Biggl( -
  \int\limits_{t_0}^{t_0+\tau} \sigma_s (\Q|q_{\max \{k:t_k < s\}}) \, ds 
  \Biggr) \, \prod_{k=1}^n \sigma_{t_k} (q_k | q_{k-1}) \,,
\end{split}
\end{equation}
with $\sigma_s(q|r)$ given by \eqref{crate} and $\sigma_s(\Q|r) :=
\sum_{q \in \Q} \sigma_s (q|r)$. Eq.~\eqref{drate} is a fact about any
jump process in continuous time with jump rates $\sigma$ (applied here
to Bell's process $Q_t$).\footnote{To get a grasp of \eqref{drate},
  begin with noting that $\sigma_s(\Q|r)$ is the total jump rate at
  time $s$ in the configuration $r$. The probability that no jump
  takes place before time $t$, if the process starts at $t_0$ in
  $q_0$, is $\exp \bigl(-\int_{t_0}^t \sigma_s(\Q|q_0) \, ds \bigr)$.
  Thus, the probability that the first jump takes place between time
  $t$ and $t + dt$ is $\exp \bigl(-\int_{t_0}^t \sigma_s(\Q|q_0) \, ds
  \bigr) \, \sigma_t (\Q|q_0) \, dt$. The probability that the
  destination of the first jump is $q_1$, given that the jump takes
  place at time $t$, is $\sigma_t (q_1|q_0) / \sigma_t (\Q|q_0)$.
  Conditional on that the first jump occurs at $t$ and leads to $q_1$,
  the distribution of the times and destinations of the further jumps
  is the same as for a process starting at time $t$ in $q_1$. Thus,
  the probability of a path $q_0, \ldots, q_n$ with the $k$-th jump
  between $t_k$ and $t_k + dt_k$ and no further jump before $t_0 +
  \tau$ is the integrand of \eqref{drate} times $dt_1 \cdots dt_n$.
  Now add (respectively integrate) the probabilities of all ways the
  process can move from $q'$ to $q$ in the time interval
  $[t_0,t_0+\tau]$, namely by means of $n$ jumps at times $t_1,
  \ldots, t_n$ with destinations $q_1, \ldots, q_n$. For a more
  detailed discussion of such probability formulas, see \cite{crex1}
  and \cite{Brei68}.}

The process $\tilde{Q}$ is not completely determined by $\Psi_0$, $U$,
and $P(\cdot)$ since $H$ is not completely determined by \eqref{UH},
even though in many cases there may be a natural choice of $H$.
For example, if $U$ has an eigenvalue $e^{-i\theta}$, then $H$ may
have as the corresponding eigenvalue any of the numbers $\frac{\theta}{\tau} +
\frac{2\pi}{\tau} k$ with $k \in \ZZZ$. More generally, for any
self-adjoint operator $S$ with spectrum contained in
$\frac{2\pi}{\tau}\ZZZ$ and commuting with $H$ (in the sense of
commuting spectral projections), $H+S$ is another solution of
\eqref{UH} for given $U$. A unique $H$ could be selected by the
additional condition that the spectrum of $H$ be contained in
$(-\frac{\pi} {\tau}, \frac{\pi} {\tau}]$.

In the particularly simple situation $|\Q| =2$, there \z{does exist a}
time-discrete analog $(\hat{Q}_t)_{t \in \tau\ZZZ}$ to Bell's process.
In this case, \z{the expression \eqref{expression} satisfies \eqref{cond} 
and thus defines a process; in fact,} the net probability current between the two
configurations $q'$ and $q$ is already determined by the distribution
\eqref{Born} and must be
\begin{equation}\label{2current}
  \sp{\Psi_t} {U^*P(q)U| \Psi_t} - \sp{\Psi_t} {P(q)| \Psi_t} \,, 
\end{equation}
since any increase or decrease can occur only by transitions from or
to the other configuration. Just as Bell's process has the smallest
jump rates compatible with the current \eqref{ccurrent} \cite{crea2A,
  schwer}, we may choose now the smallest transition probabilities
compatible with the current \eqref{2current}, which are
\begin{equation}
  \PPP_t \bigl( \hat{Q}_{t+\tau} \neq q \big| \hat{Q}_t = q \bigr) =
  \frac{\sp{\Psi_t} {(P(q)-U^*P(q)U)| \Psi_t}^+}
  {\sp{\Psi_t} {P(q)| \Psi_t}}  \,.
\end{equation}
This need not coincide with the transition \z{probability} \eqref{drate} of
$(\tilde{Q}_t)$, even though in the limit $\tau \to 0$ also
$(\hat{Q}_t)$ converges to Bell's process. \z{The same construction
can be applied to the case $|\Q|> 2$ if $U$ involves only 
pairs of configurations, i.e., if
there is a partition of $\Q$ into subsets, all of which are
either pairs or singlets, such that $P(q)UP(q') = 0$ whenever $q$
and $q'$ do not belong to the same subset.
Then \eqref{expression} still satisfies \eqref{cond} and thus
defines a process. An example of this is a quantum computing circuit, 
realized through a time
sequence of single qubit unitaries and CNOT gates. (Here, a configuration $q$
corresponds to a definite value for the computational basis observable for
each qubit.)} 

To contrast \z{the previous processes} with an example of a 
process that does not converge
to Bell's process in the limit $\tau \to 0$ but has the quantum
distribution \eqref{Born} at every time, we define the process
$(Q^*_t)_{t\in\tau\ZZZ}$ by the transition probability
\begin{equation}
  \PPP(Q^*_{t+\tau} =q|Q^*_t = q') = \sp{\Psi_{t+\tau}} {P(q)|
  \Psi_{t+\tau}}\,. 
\end{equation}
This means that for every $t$, $Q^*_t$ is independent of the past and
has the quantum distribution.  Its limit as $\tau \to 0$, in a
suitable sense, is simply the process $(\tilde{Q}^*_t)_{t\in \RRR}$
for which every $\tilde{Q}^*_t$ is independent of the past and has the
quantum distribution, a process \z{reminiscent of Bell's \cite{BellMW} 
description of} a
precise version of the ``many worlds'' interpretation of quantum
mechanics: \z{``[I]nstantaneous classical configurations [$Q$] are 
supposed to exist, and to be distributed [...] with probability $|\psi|^2$.
But no pairing of configurations at different times, as would be 
effected by the existence of trajectories, is supposed.''}

\bigskip

\noindent \textbf{Acknowledgments.}  We thank Shelly Goldstein of
Rutgers University (USA) for helpful discussions \z{and two anonymous 
referees for their comments}. R.T.~thanks the
Institut des Hautes \'Etudes Scientifiques at Bures-sur-Yvette,
France, for hospitality.  This work has been partially supported by
the European Commission through its 6th Framework Programme 
``Structuring the European Research Area'' and the contract Nr.
RITA-CT-2004-505493 for the provision of Transnational Access
implemented as Specific Support Action.

\end{document}